\documentstyle[12pt]{article}

\begin{document}
\sloppy
\thispagestyle{empty}

\mbox{}
\vspace*{\fill}
\begin{center}
{\LARGE\bf
Twist-3 in Proton Nucleon Single Spin Asymmetries
} \\
\vspace{2em}
\large
O.V. Teryaev
\\
\vspace{2em}
{\it  Bogoliubov Laboratory of Theoretical Physics,}
 \\
{\it Joint Institute for Nuclear Research,\ 141980 Dubna, Russia}\\
\end{center}
\vspace*{\fill}
\begin{abstract}
\noindent
The generation of the single spin asymmetries by the
twist 3 effects in QCD is discussed. While the short-distance
subprocesses is calculable in the Born approximation, some properties
of large distance partonic correlations may be extracted by means
of sum rules relating them to the DIS spin structure function $g_2$.
The simple model for twist 3 part of the latter is proposed. The numerical
analysis seems to favor the hypothesis about small value of the
gluonic poles contribution to the single asymmetries, while the
fermionic poles may lead to the measurable effects.

\end{abstract}
\vspace*{\fill}
\newpage
%
\section{Introduction}
\label{sect1}
The single transverse spin asymmetries are known to be one of the most subtle
effects in QCD. They should be proportional to mass scale,
and the only scale in "naive" perturbative QCD is that of the current
quark mass.
The additional suppression \cite{Kane} comes
from the fact, that single asymmetries are related to the antisymmetric
part of the density matrix. Due to its hermiticity, the imaginary
part of scattering amplitude is relevant. As a result, the spin-dependent
contribution to the hard scattering cross section starts at the one-loop
level only. More exactly, it is due to the interference of the one-loop
spin-flip amplitude and Born non-flip one.
At the same time, the Born graphs provide a leading approximation
to the spin-averaged cross section and the asymmetry is proportional to
$\alpha_s$.

The more accurate
application of perturbative QCD, including twist-$3$ effects, results
in a completely different picture.
The twist-$3$ quark--gluon correlations give rise
to the QCD single transverse asymmetries
suppressed neither by the
quark mass (it is substituted by the hadronic one $M$)
nor by $\alpha_s$ (see e.g. \cite{d,f} and refs.therein).
The collective gluon field of the polarized hadron, in which
quark is propagating, provides the latter by the mass of order
that of the hadron. This field is also the source of the phase shift
and the loop integration in the short-distance subprocess
is no more required.

\section{QCD Factorization for single asymmetries}

This qualitative picture is based \cite{d} on
a self-consistent approach to the
single asymmetries in twist--3 QCD. As a result, the
parton-like
expression was obtained. A short-distance part is calculable in
perturbative QCD with slightly modified Feynman rules. The imaginary part
is produced in the Born approximation due to the extra light--cone
integration emerging at the twist--3 level. A long-distance
contribution is described by new two-argument parton matrix elements,
the so-called quark-gluon correlators (or correlations). The latter should,
in principle, be determined experimentally from a "partonometer" process,
just like the ordinary parton distributions are determined from the deep
inelastic scattering. It is important that a mass parameter implied
by the correlators is of order of the polarized hadron mass.

Consider a hard inclusive process with a transverse polarized nucleon. The
term in the cross section proportional to twist-3 correlators can be
expressed in the form
\cite{d}
\begin{eqnarray}
d\sigma _s=\int dx_1dx_2{1\over{4}} Sp[S_{\mu }(x_1,x_2)T_{\mu }(x_1,x_2)],
\end{eqnarray}
where $S_{\mu }(x_1,x_2)$ is the coefficient function of parton subprocesses
with two quark and one gluon legs; $T_{\mu }(x_1,x_2)$ depends
on parton correlators:
\begin{eqnarray}
T_{\mu }(x_1,x_2)={M\over{2\pi}}(\hat p_1\gamma ^5s_{\mu }b_A(x_1,x_2)-
i\gamma _{\rho }\epsilon ^{\rho \mu sp_1}b_V(x_1,x_2)),
\end{eqnarray}
where $\epsilon ^{\rho \mu sp_1}=\epsilon ^{\rho \mu \alpha \beta }
s_{\alpha }p_{1\beta }$, $s_{\mu }$ is the covariant hadron polarization
vector and $M$ is the hadron mass. Two--argument distributions (correlators)
\begin{eqnarray}
b_A(x_1,x_2)={1\over{M}} \int{{d\lambda _1 d\lambda _2\over2\pi}}
e^{i\lambda _1(x_1-x_2)+i\lambda _2x_2}
\langle p_1,s|\bar \psi (0)\hat{n} \gamma ^5(D(\lambda _1)s)\psi (\lambda _2)
|p_1,s \rangle ,
\end{eqnarray}

\begin{eqnarray}
b_V(x_1,x_2)={i\over{M}} \int{{d\lambda _1 d\lambda _2\over2\pi}}
e^{i\lambda _1(x_1-x_2)+i\lambda _2x_2} \epsilon ^{\mu sp_1n}
\langle p_1,s|\bar \psi (0)\hat{n} D_{\mu}(\lambda _1)\psi (\lambda _2)
|p_1,s \rangle
\label{gefcor}
\end{eqnarray}
are real and dimensionless. They possess symmetry properties which
follow from T--invariance
\begin{eqnarray}
b_A(x_1,x_2)=b_A(x_2,x_1),\ b_V(x_1,x_2)=-b_V(x_2,x_1).
\end{eqnarray}

It is convenient to decompose these functions to the singular and regular
pieces:

\begin{eqnarray}
b_A(x_1,x_2)=\varphi_A(x_1)\delta(x_1-x_2)+b_A^r(x_2,x_1), \\ \nonumber
\ b_V(x_1,x_2)={\varphi_V(x_1)\over {x_1-x_2}}+b_V^r(x_1,x_2).
\label{decom}
\end{eqnarray}

The different form of the singular pieces is dictated by different
symmetry properties and is of different physical origin. The
"short-range" $\delta$-correlation in $b_A$ is due to the ordinary
derivative coming from the transverse momentum.

At the same time, the pole residue in $b_V$
\begin{eqnarray}
\varphi_V(x_1)= (x_1-x_2)b_V (x_1,x_2)|_{x_1=x_2}
\label{res}
\end{eqnarray}
by use of the equation of motion is just the matrix element
appearing in the "gluonic pole" contribution of
J. Qiu and G. Sterman \cite{f}.
These authors suggested such a term because of the pole
of the gluonic propagator in the partonic subprocess, producing the
imaginary phase which is necessary to have the single asymmetry. However,
it is natural to use this term for the pole of the correlation as well.

\section{Sum rules for partonic correlations}

The quark equation of
motion for each flavour allows to relate them to the "transverse"
spin-dependent quark distribution
\begin{eqnarray}
\label{sr1}
{1\over {2\pi}}
\int{ dx dy} (b_A(y,x)[\sigma(x)+\sigma(y)]-[\sigma(x)-\sigma(y)]b_V(y,x)) =
\\ \nonumber
2 \int { dx } \sigma(x) x c_T^A(x).
\end{eqnarray}

Here $\sigma(x)$ is an arbitrary test function.
The independence on the choice of the vector $n$, fixing gauge and transverse
direction
results in the relation:

\begin{eqnarray}
\label{sr2}
{1\over {2\pi}}\int{ dx dy} (b_A(y,x){{\sigma(x)-\sigma(y)}\over {x-y}} =
\int { dx } \sigma(x) (c_L^A(x)-c_T^A(x)),
\end{eqnarray}
where $c_L^A (x)$ is the most familiar distribution of the longitudinally
polarized quarks. All the integrals in (\ref{sr1},\ref{sr2}) are performed
in the regions $|x, y, x-y|\leq 1$. For ordinary parton distributions
positive arguments correspond to the quarks and the negative ones -- to
the antiquarks.
In the Born approximation the structure functions $g_1$ and $g_2$
are:
\begin{eqnarray}
\label{def}
g_1(x)+g_2(x) = {1\over 2} \sum_f e^2 (c_T(x)+c_T(-x))   \\ \nonumber
g_1(x) = {1\over 2} \sum_f e^2 (c_L(x)+c_L(-x))
\end{eqnarray}

The quark-gluon correlations in our approach combine the contributions,
related to the terms of twist 2 and 3 in the operator product expansion.
However, it appears possible to separate them just from the physical reasons.
Note that matrix elements, containing the covariant derivative,
are actually {\it not} gauge invariant. This is because the derivative
and the quark field it is acting on are taken in the different points
at the light cone. One can easily pass to the gauge-invariant form by
shifting gluon field to the point of quark field and expressing
the compensating term $A^{\mu}(\lambda_1)-A^{\mu}(\lambda_2)$
in terms of the gluon strength
(the latter is posiible because the axial gauge
is used). This contradicts the
earlier statements \cite{EFP} that transverse momentum and gluon field
are combined together in a gauge-invariant way. One may {\it separate}
these contributions, extracting the short-distance part
of the correlation (\ref{decom}),
coming from the transverse momentum and the part of gluon field
composing the covariant derivative. In this approximation
\begin{eqnarray}
b_A(x_1,x_2)=\varphi_A(x_1)\delta(x_1-x_2),
\ b_V(x_1,x_2)=0.
\label{kt}
\end{eqnarray}

The spin structure functions are then completely
determined by the transverse momentum distribution

\begin{eqnarray}
\label{gkt} g_1(x)+g_2(x) = {1\over 2} \sum_f e^2
{\varphi_A(x)-\varphi_A(-x)\over x} \\ \nonumber g_2(x) = {1\over 2} \sum_f
e^2 {d \over dx} (\varphi_A(x)-\varphi_A(-x)),
\end{eqnarray}
which may be excluded from these equations.
As a result, one get the simple differential equation,
whose obvious solution
\begin{eqnarray}
\label{wwx}
g_1(x)+g_2(x) = \int_x^1 dx {g_1(x)\over x}   \\ \nonumber
\end{eqnarray}
after calculation of the moments
\begin{eqnarray}
\label{wwn}
&&\int_0^1 x^n({n\over {n+1}}g_1(x)+g_2(x)) dx =0;
\end{eqnarray}
is nothing else than Wandzura-Wilzcek (WW)\cite{WW} sum rules.
Originally they were derived in the framework of the operator product
expansion
for the "twist-2" part of $g_2$ related to the totally symmetric operators.
Therefore, the twist-2 approximation corresponds to the taking
into account the transverse degrees of freedom of polarized quark
\cite{deut95}
(the same result was obtained recently by P. Mulders and R. Tangerman
using the different approach \cite{mul}).

One may easily calculate the gluonic corrections to these sum rule
\cite{Rule}, defining the twist-3 part $\bar g_2$, combining (\ref{sr1}),
(\ref{sr2}), (\ref{def} and (\ref{decom}) and taking into account, that
only regular piece of $b_A$ contributes:

\begin{eqnarray}
\label{wwmod}
&&\int_0^1 x^n \bar g_2(x) dx = \int_0^1 x^n({n\over {n+1}}g_1(x)+g_2(x))
dx =     \\ \nonumber && -{ 1\over \pi (n+1)} \int_{|x_1, x_2, x_1-x_2|\leq
1} d x_1 d x_2 \sum_f e^2_f [{n\over 2} b_V(x_1,x_2)(x_1^{n-1} - x_2^{n-1})
+  \\ \nonumber && b^r_A (x_1,x_2)\phi_n (x_1,x_2)],
\ \ \phi_n (x,y)={{x^n-y^n}\over {x-y}}-{n\over 2}(x^{n-1}-y^{n-1}),\ \
n=0,2...
\end{eqnarray}

Let us consider the important particular cases. For the analysis of the
famous Burkhardt-Cottingham \cite{BC} sum rule ($n=0$)
one may use the expression for $g_2$, generated by
the (regularized) $\delta$-function as a test function
in (\ref{sr2})
\begin{equation}
g_2(x) ={1\over {2\pi}}\int_{|x,y,x-y|\leq 1}dxdy\
{b^A(x,y)+b^A(y,x)\over {x-y}}.
\end{equation}

One of the  possibilities for BC sum rule violation is
the {\it long-range} singularity $\delta(x)$ \cite{Jaf}.If $g_2$ contains
such a term proportional to
$\delta(x)$, which can never be observed experimentally, it will give a
non-zero contribution to the integral and therefore will
violate the BC sum rule. Note that such a situation was
first noticed by Ahmed and Ross in their pioneer paper on  spin
effects in QCD
\cite{Ah}, where they also mentioned a similarity with the Schwinger
 term sum rule for the longitudinal structure function.
However, to obtain such a behavior, one also needs to have a
singularity  in $b^A$  e.g.
\begin{equation}
b^A(x,y)=\delta(x) \phi(x,y)+\delta(y) \phi(y,x),
\end{equation}
where $\phi$ is regular function. Such a singularity should result
in  meaningless infinite fermionic poles (when $x=0$) contribution to
single asymmetries.

While these arguments allow one to impose some restrictions for $g_2$
starting from single asymmetries, the most interesting for our
purposes is the 'opposite' problem: what can we learn about
twist 3 single asymmetries having some information about twist 3
contribution to $g_2$.

\section{Quantitative estimates of single asymmetries and model for
twist-3 part of $g_2$}

The first numerical estimate was performed a couple of years ago
\cite{shafer}. The starting point was the
value of $\int_0^1 dx x^2 g_2 (x)$, calculated in the framework of
QCD sum rule method \cite{Braun}.
Using the model assumptions about the constance of chromomagnetic field
inside the nucleon, the authors estimated quantitatively the gluonic
poles contribution to the single asymmetry of direct photon production.

To understand this problem better it instructive to consider (\ref{wwmod})
for n=2 \cite{Rule}.

\begin{eqnarray}
\label{n2}
&&\int_0^1 x^2 \hat g_2(x) dx =     \\ \nonumber
&& -{ 1\over 3 \pi } \int_{|x_1, x_2, x_1-x_2|\leq 1} d x_1 d x_2
\sum_f e^2_f b_V(x_1,x_2)(x_1- x_2).
\end{eqnarray}

Note that only {\it vector} correlation contribute to this first nonzero
moment.  This fact, known for many years, is especially clear from the
non-local OPE \cite{grt} in the co-ordinate space:

\begin{eqnarray}
\label{p1.4}
\lefteqn{
 \{T{\bar\psi}\gamma^\mu \psi(x)
   {\bar\psi}\gamma^\nu \psi(y)\}_{[\mu \nu]} =
    \epsilon^{\mu\nu\rho\sigma} \{i\partial_\rho^{x} D^c(x,y)
    {\bar\psi}(x)\gamma_\sigma\gamma_5 U(x,y)
    \psi(y) } \\
 &- &g D^c(x,y) {\bar\psi}(x)\gamma_\sigma\gamma_5
    \int_0^1 dt(t-\frac{1}{2})(x-y)^\lambda U(x,z)  F_{\rho\lambda}(z)
     U(z,y) \psi(y)\nonumber \\
  &-&i g D^c(x,y) {\bar\psi}(x)\gamma_\sigma
    \int_0^1 dt\frac{1}{2} U(x,z)  {\tilde F}_{\rho\lambda}(z(t))
    (x-y)^\lambda
     U(z,y) \psi(y) \}.
     \nonumber
\end{eqnarray}

The term ${\bar\psi}(0)\gamma_\sigma\gamma_5 F_{\rho\lambda}(0) \psi(0)$
contributing to the second moment of $g_2$
contains the factor $\int_0^1 dt(t-\frac{1}{2})=0$

The sum rule (\ref{n2}) is especially simple when one neglect the regular
piece of $b_V$:

\begin{eqnarray}
\label{n2pole}
\int_0^1 x^2 \bar g_2(x) dx =
-{ 1\over 3 \pi } \int_{|x_1, x_2, x_1-x_2|\leq 1} d x_1 d x_2
\sum_f e^2_f \varphi_V(x_1).
\end{eqnarray}

Performing the integration over $x_2$ one get:

\begin{eqnarray}
\label{n2polem}
\int_0^1 x^2 \bar g_2(x) dx =
-{ 1\over 3 \pi } \int_{-1}^1 d x_1 \sum_f e^2_f \varphi_V(x_1)(2-|x_1|).
\end{eqnarray}

The contribution of the first term in the bracket coincides, up to a
constant of order 1, with the sum rule derived \cite{shafer} by use of
the model assumption about the constant collective gluon field. This
is quite natural, because the gluonic pole in the correlation is produced
by the integration over large distances at the light cone, corresponding
to the quasiclassical collective fields. The second term, effectively
increasing the correlation for given value of $\bar g_2$, is produced
by the shape of integration region. The latter is related to the analytical
properties of the correlations and is therefore a pure quantum effect.

The second moment of $\bar g_2$ was calculated, as it was mentioned above,
in the framework of
QCD sum rules\cite{Braun}. These calculation was checked recently
\cite {shafer2}
by use of the another interpolating nucleon currents, explicitly accounting
for the gluonic degrees of freedom. The results are compatible:
this moment is small (about $10^{-2}$) and negative for the proton case,
while for the neutron it is about of order of magnitude larger.
Consequently, the single asymmetry of direct photon production is claimed
to be small and strongly flavour-dependent.

One should note, however, that only gluonic poles contribution calculated by
J. Qiu and G. Sterman was taken into account. At the same time,
the contribution of fermionic poles
(when imaginary part is produced by the pole of quark
propagator) is related to the correlation $b_A$, which
is not manifested at all
for the second moment of $g_2$.

While the fermionic poles were
suggested originally as a source of single asymmetries \cite{d},
the gluonic poles were discovered later \cite{f} and claimed the
dominant  contribution (because it is more easy to emit
the zero-momentum gluon than zero momentum quark; note, however,
that such emission is known to produce
the IR divergencies and may therefore cancel somehow).

Now, when QCD sum rules tell us that gluonic poles are, contrary to these
expectations, small, there are two opportunities. First (seems to be
now silently accepted) is to believe that the contributions of
fermionic poles are even smaller. Second, which we study here,
is to come back to fermionic poles and investigate them more seriously.

To do this one may, say, repeat the QCD SR calculation  substituting
$\tilde F \to  F, \gamma_\sigma \to \gamma_\sigma \gamma_5$. This operator,
as it was already explained, does not contribute to $g_2$. However, its
non-local generalization is directly related to $b_A$. If one observe
its magnitude larger enough than that of vector operator, this would be
the strong support that only the gluonic poles are small. The same
may be probably performed by modifying the existing lattice \cite{lat}
calculation
of the second moment of $\bar g_2$.

As we do not know the value of this
matrix element at the moment, let us try to perform at least the model
estimates. To do this, another sum rule, relating the {\it first}
moments of $g_1$ and $g_2$, is extremely useful. It is known for several
years \cite{Rule}, although the role of valence quarks was emphasized
only recently \cite{OT94}.

The applicability of the correlations to the single asymmetries is possible
because they are the (generalized) distribution functions on the light cone,
accumulating the large-distance dynamics. This allows one to apply
(\ref{sr1},\ref{sr2}) with the {\it odd} test function and produce
sum rules, which can never appear in the framework of the operator product
expansion, because they
corresponds to the "wrong" moments (the positive parity of two photons
pick up the even combinations only). In these cases the combinations
$c_{L,T}(x)-c_{L,T}(-x)$  appear. They are just the {\it valence} quark
contributions.
As a result, the sum rules (\ref{wwmod}) are valid for $n=1,3...$
with the $g_1, g_2$ in the l.h.s. being replaced by their valence
parts. The case of $n=1$ is especially interesting, because the contribution
of the correlations in the r.h.s. appears to be equal to zero.
\begin{eqnarray}
\label{SR}
\int_0^1 dx x (g_1^{val} (x) + 2 g_2^{val} (x)) = 0
\end{eqnarray}

This sum rule allows one to relate such a "model" entities as valence
quark starting from the QCD correlations.
The situation here is in some sense opposite to, say, Gottfried sum rule.
Instead of "model" derivation for measurable functions, we have
(more or less) rigorous QCD derivation for "model" functions.
Note that this sum rule is
not affected by the possible $1/x^2$ singularity, spoiling the validity
of BC sum rule \cite{AEL}.

The valence contribution to $g_2$ may be measured, in principle, by
considering the semi-inclusive asymmetries in DIS on transversely polarized
target, in complete analogy with the studies of longitudinal
valence distributions at CERN \cite{semi}. Although the transverse
polarization is, generally speaking, the non-probabilistic twist 3 effect,
the parton-like formula (\ref{def}) for DIS allows one to extract the valence
parts of $c_T$ (and, by use of (\ref{sr1}), of the correlations),
applying the same formulas as for the longitudinal polarization.
One should just change the longitudinal semi-inclusive asymmetries
for the transverse ones, and $c_L$ to $c_T$.

The alternative derivation of this sum rule \cite{AEL} leads
to its validity for {\it total} structure functions. One may guess that
it is valid for quark and antiquarks separately. The other opportunity
is \cite{ELT} that the quark-gluon correlations are negligible, if
the signs of $x_1$ and $x_2$ are opposite (i.e., the correlation
of gluon and quark-antiquark pair is considered). It is interesting,
that in this case one should change $2-|x_1| \to 1$ in the sum rule for
the second moment (\ref{n2pole}) and get the expression completely analogous
to its model derivation \cite{shafer}. Anyway, let us neglect in the
following the difference between valence and total contribution to
$g_2$.

As a result,
the WW sum rules appear to be exact (in the leading twist 3 approximation)
for the {\it two} moments. Consequently,
 the deviation of $g_2$ from its WW value
(twist 3 piece) has these moments equal to zero. This function
should change sign at least two times inside the region $(0,1)$.
The similar behavior was predicted for the singlet piece of the
tensor spin structure function\cite{ten}.

Let us incorporate this property to the following
parametrization of the twist 3 contribution, where, as usual, the Regge
behavior for small $x$ and the phase space for $x \sim 1$ are manifested:

\begin{eqnarray}
\label{param}
\bar g_2(x)=C x^{-\alpha}(x-a_1)(x-a_2)(1-x)^{\beta}
\end{eqnarray}

Imposing the conditions
\begin{eqnarray}
\label{cond}
\int_0^1 dx \bar g_2 (x) =\int_0^1 dx x \bar g_2 (x) = 0
\end{eqnarray}
allows us to find the crossing points $a_1$ and $a_2$ in terms of
$\alpha$ and $\beta$. At the final stage, we shall calculate the
second moment of $g_2$.

We performed these calculation for $\beta=4$ and $\alpha$
changing in the
interval [0,1). As one should not take this parametrization too seriously,
the general features, not very sensitive to the actual values of parameters,
are most important:

i) The function has the negative minimum at $x \sim 0.12 \pm 0.05$ and the
positive maximum at $x \sim 0.6 \pm 0.05$.

ii) The second moment is rather small and negative -- about $-10^{-3} C$.

iii) The ratio of the value of function in the minimum to the  value of the
second moment is changing in the interval $-(70-130)$,
while for the maximum it is $+(10-20)$.

One may expect, that the small value of the second moment, predicted
by QCD sum rules, is reflecting the strong oscillations rather than
small normalization. This is also compatible with the dominance of axial
correlation, absent for the second moment but manifested in another moments
and in the fermionic poles contributions to single asymmetries.

Taking into account the negative QCD SR value, one should expect
the maximal value of $\bar g_2$ at $x \sim 0.1$ to be {\it positive}.

\begin{eqnarray}
\label{SRg23}
\bar g_2^p (x=0.12 \pm 0.05) = 0.16 \pm 0.10;  \\ \nonumber
\bar g_2^n (x=0.12 \pm 0.05) = 0.8 \pm 0.4;
\end{eqnarray}

While the preliminary data of E143 experiment reported in Morionde
were negative in this region (supporting the positive value of the
second moment), the more recent
data \cite{rock} are indicating (for one of two spectrometers) the positive
values compatible with (\ref{SRg23}).  However, they are still compatible
with zero as well, and the more precise HERMES data are most welcomed.

\section{Short-distance subprocesses}

Let us discuss briefly the calculation of the short-distance QCD
subprocess. It is determined by Born graph
(the imaginary part is produced by the pole of the propagator while the
integration over light-cone variable is performed) and the computation is
straightforward. Note that a number of fermionic poles
contributions to quark-gluon subprocesses is already calculated \cite{KT}.
This offer a possibility not to limit the future experimental studies by the
asymmetries of the direct photon production.

I would like to stress the special role of the single asymmetries of jet
production. This asymmetry is just
\begin{equation}
A={{\sigma_L-\sigma_R}\over {\sigma_L+\sigma_R}},
\end{equation}
where $\sigma_L$ ($\sigma_R$) is just the cross-section for jets
produced to the left (right) of the normal to scattering plane.
This does not require the full analysis of jet structure, like
for handedness studies \cite{ladin}, and only the trust direction is
necessary. At the same time, the cross-sections are much larger than
in the photon case. In the proposed experiment $HERA-\vec N$, when
800 GeV proton beam would scatter on the polarized nucleon target,
the expected statistical error for the jet left-right asymmetry is
less than $10^{-3}$. This would allow one to measure the asymmetry
of order $10^{-2}$, expected for "pessimistic" estimates
based on the values (\ref{SRg23}),

From the theoretical point of view, these effects are
just the asymmetries of quark and gluon production. They are insensitive
to the fragmentation effects and provide as clear test of QCD, as
a photon asymmetry.

Another interesting process, for which calculations are already performed,
is the dilepton left-right asymmetry. The dilepton mass variation
may be used to probe the quark-gluon correlation in various regions of
the ligh-cone momentum fractions. At the same time, the background
problem can be solved more easy in this case.

The quark gluon subprocesses contribute also to the single asymmetries
of pion production. It is interesting, that the well-known problem
of flavour dependence can be naturally solved, if the contribution
of the fermionic poles  is much larger than that of the gluonic poles.

The non-existence of the gluonic poles leads to the significant
numerical difference between the single asymmetries of the gluon

\begin{eqnarray}
\label{f}
\nonumber
A_{gg}=& &\!\!\!\!\!\!\!\!\!{b_A(0,x)-b_V(0,x)\over{f(x)}}{Mp_T\over{m_T^2}}
\nonumber \\
& &\ \ \ \ \ \ \ \ \ \times
{(C_F-C_A/2)(1-x_F)\over{C_F(1+x_F^2)}}
{((x_F^4+1)C_A/2-C_Fx_F(1-x_F)^2)\over{(x_F(C_F-C_A/2)-(1+x_F^2)C_F/2)}},
\nonumber
\end{eqnarray}

and quark
\newpage
\begin{eqnarray}
\label{g}
A_{gq}={b_A(0,x)-b_V(0,x)\over{f(x)}}{Mp_T\over{m_T^2}}\hspace{8cm}
\nonumber \\
\times{1\over C_F(C_F(x_F^2(1-x_F)+x_F^4/2)+C_A/2(2-4x_F+3x_F^2-x_F^3))}
\nonumber \\
\times\left( C_F^2(x_F^2(x_F-1))-C_FC_A/2(x_F^3-x_F^2-x_F)+ \right.
\hspace{4cm} \nonumber\\
\left. +C_A^2/4(x_F^5-5x_F^4+11x_F^3-14x_F^2+9x_F-4)\right)
\nonumber
\end{eqnarray}
production on the transverse polarized nucleon by the gluon from the
unpolarized one. Namely, the latter
is several times larger\cite{EKT}. One may expect, that the pion single
asymmetry is due to the pions resulting from the quark (not gluon!)
fragmentation. As a result, the $\pi ^+$ meson asymmetry is related
to the correlation of gluon and $u$-quark, while $\pi ^-$ meson asymmetry
is related to the correlation of the gluon and $d$-quark; $\pi ^0$
can be produced by $u$- and
$d$-quarks with equal probability.

Making use of sum rules relating the correlations to the "ordinary"
quark distributions (\ref{sr1}), valid for each flavour separately,
one can easily get
\begin{eqnarray}
A_{\pi ^+} \sim {\Delta u\over u}, \ \ \ \ A_{\pi ^-} \sim {\Delta d\over d},
\ \ \ \ \ A_{\pi ^0} \sim {{\Delta u+\Delta d}\over{u+d}}.\nonumber
\end{eqnarray}

Note that mirror
asymmetries for $A_{\pi ^+}$ and $A_{\pi ^-}$
and $A_{\pi ^0}=1/3A_{\pi ^+}$
can be obtained if $u=2d$
and $\Delta u=-2\Delta d$, what is not far from
experimental observation.

\section{Conclusions}

The earlier quantitative estimates of the twist-3 single asymmetries
are dealing, in fact, only with the contributions of the gluonic poles.
their smallness therefore should be interpreted as the argument against
the original hypothesis of J. Qiu and G. Sterman about the domination
of gluonic poles.
The relative strength of fermionic and gluonic poles
may be checked by the calculation of "dual" quark-gluon matrix element
using QCD sum rules or lattice simulations.

As the solution of this problem by use of nonperturbative
methods was not found yet,
the models of twist 3 effects are especially important.
We are suggesting the simple model for the twist 3 part of $g_2$ structure
function. It incorporates the Regge and phase space behavior,
Burkhardt-Cottingham sum rule as well as another sum rule relating the
first moments of $g_1$ and $g_2$.  It is found, that the smallness of the
second moment is the result of the strong oscillations required by sum
rules, while its absolute value may be fairly large at the region of $x
\sim 0.1$.
The values of twist-3 part of $g_2$ in this region are predicted
in the proton (about 0.15) and neutron (about 0.8)
cases, using the QCD sum
rules calculations for the second moment.
Such a behavior is supporting the domination of the fermionic
poles over the gluonic ones, because it is the second moment, where the
correlation producing the fermionic poles does not contribute.

The existing perturbative calculations allow one to study the various
single asymmetries in the proton-nucleon scattering at 800 GeV
(proposed experiment $HERA-\vec N$). The left-right asymmetries
of jets and dileptons are especially interesting, as providing
good statistics and offering the opportunity of clear tests of twist 3 QCD.

{\it Acknowledgements.

I am indebted to J.~Bl\"umlein,
A.V.~Efremov, V.M.~Korotkiyan,
E.~Leader, P.~Mulders,
W.-D.~Nowak, A.~Sch\"afer and R. Tangerman for valuable discussions.
I would like to thank P. S\"oding for warm hospitality at DESY-Zeuthen,
where the substantial part of this investigation was performed.
This work was possible due to the support of International Science
Foundation (Grant RFE300), Russian Foundation for Fundamental Research
(Grant 93-02-3811)
and Government of Land Brandenburg}.

\end{document}